\documentclass[11pt, a4paper]{article}

\usepackage{amsmath}
\usepackage{amsfonts}
\usepackage{amssymb}
\usepackage{graphicx, rotating}
\usepackage{epstopdf}
\usepackage{epsfig}
\usepackage{latexsym}
\usepackage{graphicx}
\usepackage{color}
\usepackage{amsmath,bm,amssymb}
\usepackage{cite}
\usepackage{slashed}
\usepackage{hyperref}
\usepackage[numbers,sort&compress]{natbib}
\hypersetup{colorlinks, citecolor=bluscuro, linkcolor=black, urlcolor=bluscuro}
\definecolor{rossos}{cmyk}{0,1,1,0.55}
\definecolor{bluscuro}{rgb}{0.15, 0.2, .85}
\definecolor{bluchiaro}{cmyk}{1,.3,0.,0.1}

\newcommand{\be}{\begin{equation}}
\newcommand{\ee}{\end{equation}}
\newcommand{\bea}{\begin{eqnarray}}
\newcommand{\eea}{\end{eqnarray}}

\def\simlt{\stackrel{<}{{}_\sim}}
\def\simgt{\stackrel{>}{{}_\sim}}

\newcommand{\GeV}{\,\mathrm{GeV}}

\begin{document}

\begin{titlepage}
\begin{flushright}
IFT-UAM/CSIC-19-109
\end{flushright}
\begin{center} ~~\\
\vspace{0.5cm} 
\Large {\bf\Large UV completion of an axial, leptophobic, $Z'$} 
\vspace*{1.5cm}

\normalsize{
  {\bf  J.~A.~Casas$^a$\footnote[1]{alberto.casas@uam.es},
    M.~Chakraborti$^a$\footnote[2]{mani.chakraborti@gmail.com},
J.~Quilis$^a$\footnote[3]{javier.quilis@csic.es}} \\

\smallskip  \medskip
$^a$\emph{Instituto de F\'\i sica Te\'orica, IFT-UAM/CSIC,}\\
\emph{U.A.M., Cantoblanco, 28049 Madrid, Spain}}
%\\
%$^c${\it Department of Physics, University of Notre Dame,}\\
%{\it Notre Dame, IN 46556, USA}}\\
%$^d$\emph{ARC Centre of Excellence for Particle Physics at the Terascale \\
%School of Physics, The University of Melbourne, Victoria 3010, Australia}

\medskip
%Version as of \today

\vskip0.6in 

\end{center}

%%%%%%%%%%%%%%%%%%%%%%%%%%
%%      Abstract        %%
%%%%%%%%%%%%%%%%%%%%%%%%%%
\centerline{\large\bf  Abstract}
\vspace{.5cm}
\noindent

The $Z'$-portal is one of most popular and well-explored scenarios of dark matter (DM).
To avoid the strong constraints coming from dilepton resonance
searches at the LHC and direct detection of DM, it is usually required that
in addition to being leptophobic, the $Z'$ is axially coupled to either the
(fermionic) DM or the standard model (SM) quarks.
The first possibility has been extensively studied both in the context of
simplified model and UV complete scenarios. However, the studies on the second possibiliy
are largely confined to simplified models only. Here, we construct the minimal UV completion of these
models satisfying both the criteria of leptophobia and purely axial $Z'-$quark
coupling. The anomaly cancellation conditions demand
highly non-trivial structures, not only in the dark sector, but also in the Higgs sector. We also discuss the main phenomenological implications of the UV completion.

\vspace*{2mm}
\end{titlepage}

\section{Introduction}
One of the most popular scenarios for dark matter (DM) consists of a
Standard Model (SM) fermionic singlet, $\chi$ (the DM particle) coupled
to SM fields via a massive $Z'$ gauge boson
(see e.g. \cite{Langacker:1984dc,Langacker:2008yv,FileviezPerez:2010gw,Frandsen:2011cg,
    Duerr:2013dza,Duerr:2013lka,Alves:2013tqa,Arcadi:2013qia,Lebedev:2014bba,Duerr:2014wra,
    Perez:2015rza,Duerr:2015vna,Kahlhoefer:2015bea,Jacques:2016dqz,Fairbairn:2016iuf,Arcadi:2017hfi,
    Perez:2014qfa,Ohmer:2015lxa,Ismail:2016tod,Okada:2016tci,Okada:2017dqs,Bandyopadhyay:2018cwu,Pandey:2018wvh,Okada:2018tgy,Das:2019pua,Blanco:2019hah}).
Typically, the most severe constraints on these kinds of models come from  di-lepton
production at the LHC \cite{Aaboud:2017buh,Sirunyan:2018exx} and from DM direct-detection
experiments \cite{Aprile:2018dbl}. Concerning the first ones,
a usual strategy is to consider leptophobic models, so that the $Z'$ just couples to quarks
in the SM sector. Similarly, spin-independent direct-detection cross-section is drastically
suppressed if the $Z'$ has axial couplings either to the DM particle or to
the quarks (or both) \cite{Lebedev:2014bba,Hooper:2014fda,Kahlhoefer:2015bea,Ismail:2016tod,Ellis:2018xal,Bagnaschi:2019djj}.

Usually, the phenomenological analyses have been done
in the context of simplified DM models, where the DM particle and the $Z'$
mediator are the only extra fields (see e.g. \cite{Buchmueller:2014yoa}).
The corresponding parameter-space is then 
spanned by the $Z'-$mass, its coupling to the DM particle and the coupling(s) to the SM fields. 

However, the above view becomes over-simplified when one takes into account
theoretical constraints, in particular those coming from the requirement of anomaly cancellation.
In this sense, there have been a number of studies exploring possible ultraviolate (UV) completions
of the leptophobic $Z'$ scenario  when the $Z'$ boson has vectorial coupling to  quarks
and (preferably) axial coupling to the DM particle {\cite{Ellis:2018xal,Caron:2018yzp,ElHedri:2018cdm}.
In that case, the most important conclusion
that emerges is that the dark sector (DS) has to be enlarged beyond the most simplified picture.
More precisely, the minimal DS consists of the DM particle, $\chi_{L, R}$, a $SU(2)$ doublet,
$\psi_{L,R}$, and a $SU(2)$ singlet, $\eta_{L,R}$. On the other hand,
the charges of these fields under the extra $U(1)$ are fixed by
anomaly cancellation, thus  reducing the effective parameter-space;
although there appear new parameters related to the extra stuff in the DS.

The complementary scenario, when the leptophobic $Z'$ boson has purely
axial coupling to the SM quarks and vectorial/axial vector coupling to the DM particle has been often considered in phenomenological analyses (usually in the context of simplified models) \cite{Lebedev:2014bba,Hooper:2014fda,Ismail:2016tod,Bagnaschi:2019djj}; but its possible UV completions remain mostly unexplored, except for Ref.~\cite{Ismail:2016tod}. 
The main goal of this paper is to determine the form of the minimal DS for this scenario, consistent with anomaly cancellation, and the complete set of consistent assignments of ordinary and extra hypercharges to the various fields. As for the vectorial case, the DS must be extended with respect to the usual assumptions in simplified models; actually the extension is larger than for the vectorial case. 

In addition, we show that for any, minimal or not, UV completion, the consistency of the scenario requires the Higgs sector to contain at least three Higgs doublets .

  In Sec.\ref{sec:higgssector} we outline the structure of the Higgs sector as required by
  the conditions of leptophobia and axial couplings to quarks. In Sec.\ref{sec:anomaly}
we derive the constraints on the particle content of the models coming from
the anomaly cancellation conditions. In Sec.\ref{sec:minimal}, we present the minimal
scenario consistent with all the requirements, giving a complete account of the possible assignments of charges to the various fields. Sec.\ref{sec:pheno} contains a brief discussion on the main phenomenological features of the minimal UV completion. Finally, we conclude in Sec.\ref{sec:conc}.

\section{Constraints in the Higgs sector}
\label{sec:higgssector}
Let us start by showing that a leptophobic $Z'$ axially coupled to quarks
requires a Higgs sector consisting of, at least, three Higgs doublets.
If the Higgs sector contains just one Higgs (as in the SM), then the
invariance under the extra gauge factor, $U(1)_{Y'}$, of the fermionic Yukawa couplings
\be
y^e_i \bar L_i H e_i,\ \ \ \  y^u_i \bar Q_i \bar H u_i,\ \ \ \  y^d_i \bar Q_i H d_i
\ee
($y_i$ are the Yukawa coupling constants, with $i$ a family index), forces
the $Y'$-charge of the Higgs to vanish, $Y'_H=0$, in order to satisfy the
leptophobia assumption ($Y'_L=Y'_e=0$). On the other hand, since
$Y'_Q =-Y'_u=-Y'_d$ (axial-coupling assumption), the invariance of the
above hadronic Yukawa couplings implies $Y'_{Q_i}=Y'_{u_i}=Y'_{d_i}=0$,
so there is no coupling at all to quarks.

For a two-Higgs doublet (THDM) model things are similar.
Suppose that in the THDM under consideration $u-$ and $d-$quarks
couple to the same Higgs, say $H_1$. This is the case of Type I and lepton-specific
THDMs \cite{Branco:2011iw}. Then, the invariance of the hadronic Yukawa couplings,
\be
y^u_i \bar Q_i \bar H_1 u_i,\ \ \ \  y^d_i \bar Q_i H_1 d_i\ ,
\ee
plus the axial requirement ($Y'_Q =-Y'_u=-Y'_d$) imply $Y'_{H_1}=Y'_{Q_i}=Y'_{u_i}=Y'_{d_i}=0$. 

Suppose now that $d-$quarks couple to a Higgs doublet, say $H_1$, different to that of $u-$quarks,
say $H_2$. This is the case of Type II and flipped THDMs \cite{Branco:2011iw}.
Since one of the two Higgses must couple to leptons, either $Y'_{H_1}=0$ or $Y'_{H_2}=0$.
Then the axial condition plus the invariance of the hadronic Yukawa couplings,
\be
y^u_i \bar Q_i H_2 u_i,\ \ \ \  y^d_i \bar Q_i H_1 d_i\ ,
\ee
imply $Y'_{H_1}-Y'_{H_2}=0$, and thus finally all the $Y'$ hypercharges are vanishing in the SM sector.
Consequently, the minimal number of Higgses to implement a leptophobic $Z'$ with axial couplings to
quarks is three, say $H_u$, $H_d$, $H_l$, each one of them coupled specifically to $u-$quarks,
$d-$quarks and leptons respectively. This conclusion is completely general, independently of the UV completion of the model.

\section{Constraints from anomaly cancellation}
\label{sec:anomaly}
Let us now obtain the conditions that anomaly cancellation imposes
on the dark sector. From now on we will assume that the $U(1)_{Y'}$
group is flavour-blind. This is a sensible assumption since, otherwise,
a not-too-heavy $Z'$ would naturally lead to dangerous FCNC.
On top of that, if the $U(1)_{Y'}$ charges of $u-$ and $d-$quarks
are family-dependent, the off-diagonal terms of the corresponding
Yukawa  matrix (necessary to reproduce the observed CKM matrix)
would be forbidden unless they arise from the coupling of the
quarks to extra Higgs-doublets. This would lead to further
extensions of the Higgs sector. Besides, 
the mass-eigenstates of the quarks would not
have well-defined $U(1)_{Y'}$ charges, thus spoiling their axial coupling to the $Z'$.

Therefore, the three generations of the SM fermions transform under
the gauge group, $SU(3)_C \times SU(2)_L\times U(1)_{Y}\times U(1)_{Y'}$, as 
\bea
&&Q\ \ (\ 3,\ \ \ 2,\ \ \ \frac{1}{6},\ \ \ Y'_Q\ ) ,
\nonumber\\
&&u_R\ (\ 3,\ \ \ 1,\ \ \ \frac{2}{3},\ -Y'_Q\ ) ,
\nonumber\\
&&d_R\ (\ 3,\ \ \ 1,\ -\frac{1}{3},\  -Y'_Q\  ) ,
\nonumber\\
&&L\ \ (\ 1,\ \ \ 2,\ \  -\frac{1}{2},\ \ \ \ \, 0\ ) ,
\nonumber\\
&&e_R\ (\ 1,\ \ \ 1,\ \ -1,\ \ \ \ \ 0\  ) .
%\nonumber\\
%&&\eta_R\ (\ 1,\ \ -1,\ -\frac{3}{2}\ ),
%\nonumber\\
%&& S\ \  (\ 1,\ \ \ \,0,\ \ \ -3\ ).\phantom{\frac{1}{2}}
\label{SMcharges}
\eea
In addition, we will often take $Y'_Q=1$ with no loss of generality (it is a normalization factor for the extra hypercharge).

The first consequence of these axial $U(1)_{Y'}$ charges of quarks is that there are six new anomalies to be considered:
\begin{equation}
\begin{split}
&SU(3)_C^2 \times U(1)_{Y^\prime}\\
&SU(2)_L^2 \times U(1)_{Y^\prime}\\
&U(1)_Y^2 \times U(1)_{Y^\prime}\\
&U(1)_Y \times U(1)_{Y^\prime}^2\\
&U(1)_{Y^\prime}\\
&U(1)_{Y^\prime}^3\ .
\end{split}
\label {anomalies}
\end{equation}
Out of them, only the fourth one is cancelled inside the SM sector.
Hence, the existence of a dark sector (DS) to implement anomaly cancellation is compulsory. 
Since we are interested in the minimal DS able to do that job, all the DS fermions, say $f$, must be vectorial under the
ordinary hypercharge, $U(1)_Y$, i.e. $Y_{f_L}=Y_{f_R}\equiv Y_f$, so that the four SM
anomalies, $SU(3)_C^2 \times U(1)_{Y}$, $SU(2)_L^2 \times U(1)_{Y}$, $U(1)_Y^3$ and $U(1)_Y$,
are kept vanishing. Otherwise, the DS has to be further increased (this holds for all the scenarios analyzed in the paper).

In order to play the role of the DM particle, the DS must contain a neutral particle, singlet under $SU(3)_C \times U(1)_{em}$. The simplest possibility is a fermion, $\chi_{L,R}$, singlet under the the whole SM gauge group, $SU(3)_C\times SU(2)_L\times U(1)_Y$. Then, additional fields in the DS are needed in order to cancel the anomalies of Eq.(\ref{anomalies}); in particular those associated to $SU(3)_C^2 \times U(1)_{Y^\prime}$ and $SU(2)_L^2 \times U(1)_{Y^\prime}$, which require non-trivial representations under $SU(3)_C \times SU(2)_L$. Thus the cheapest option (if viable) would be to use one extra particle, say  $\Gamma_{L, R}$, transforming as $(3,2)$. However, the corresponding equations for anomaly-cancellation read

\begin{equation}
\begin{split}
& 12 Y^\prime_Q+2(Y^\prime_{\Gamma_L}-Y^\prime_{\Gamma_R})=0, \\
& 9 Y^\prime_Q+3(Y^\prime_{\Gamma_L}-Y^\prime_{\Gamma_R})=0,
\end{split}
\end{equation}
which are compatible only if $Y^\prime_Q=0$. 

Consequently, we have to incorporate additional fields to the DS. The most economical alternative is to consider,
beside the DM particle $\chi_{L,R}$, one $SU(3)_C$ triplet, $\Phi_{L,R}$, and one $SU(2)_L$ doublet, $\psi_{L, R}$. Hence, the DS spectrum reads
\bea
&&\chi_L\ \ (\ 1,\ \ \ \,1,\ \ \ 0,\ \ \ Y'_{\chi_L}\ ) ,
\nonumber\\
&&\chi_R\ \ (\ 1,\ \ \ \,1,\ \ \ 0,\ \ \ Y'_{\chi_R}\ ) ,
\nonumber\\
&&\Phi_L\ \ (\ 3,\ \ \ \,1,\ \ \ Y_{\Phi},\  Y'_{\Phi_L}\  ) ,
\nonumber\\
&&\Phi_R\ \ (\ 3,\ \ \ \,1,\ \ \ Y_{\Phi},\  Y'_{\Phi_R}\  ) ,
\nonumber\\
&&\psi_L\ \ (\ 1,\ \ \ \,2,\ \ \ \ Y_{\psi},\  Y'_{\psi_L}\  ) ,
\nonumber\\
&&\psi_R\ \ (\ 1,\ \ \ \,2,\ \ \ \ Y_{\psi},\  Y'_{\psi_R}\  ) .
%\nonumber\\
%&&\eta_R\ (\ 1,\ \ -1,\ -\frac{3}{2}\ ),
%\nonumber\\
%&& S\ \  (\ 1,\ \ \ \,0,\ \ \ -3\ ).\phantom{\frac{1}{2}}
\label{SpecCase1}
\eea
The corresponding cancellation conditions for the six anomalies of Eq.(\ref{anomalies}) are given by Eqs.~(\ref{case1}) in the Appendix. We show that they only have non-trivial solution ($Y'_Q\neq 0$)  if $Y_\psi=\pm1/2$, $Y_{\Phi}=\pm 1/6$.
Solving the complete set of equations (\ref{case1}) we find the 8 possible
assignments of charges for the DS, which are presented in Table \ref{table:SolutionsCase1}.

To summarize, the DS of Eq.(\ref{SpecCase1}) with the charges of Table \ref{table:SolutionsCase1} represents
the most economical UV completion of a leptophobic $Z'$ with axial couplings
to quarks. Nevertheless, the fact that the dark quarks ($\Phi$) have
electric charge $Q^{\rm el}_\Phi=\pm1/6$ strongly suggests the existence of
stable baryons with fractional electric charge, e.g. $\pm 1/2$, which
would be cosmologically disastrous \cite{Chang:1996vw,Munoz:2001yj}.
Hence, we consider this possibility unrealistic.

\vspace{0.3cm}
%\noindent
There is another, in principle equally economical, alternative for the DS
when the DM particle is the neutral component of the doublet, $\psi$.
This requires $Y_\psi=\pm 1/2$ from the beginning. 
Then, one could try to satisfy the anomaly-cancellation
conditions just with the addition of a $SU(3)_C$ triplet,
$\Phi_{L,R}$ (to cancel the color anomaly) plus a singlet field,
$\eta_{L,R}$. The corresponding spectrum of the DS is similar to the previous case:
\bea
&&\psi_L\ (\ 1,\ 2,\ Y_{\psi},\  Y'_{\psi_L} ) ,
\nonumber\\
&&\psi_R\ (\ 1,\ 2,\ Y_{\psi},\  Y'_{\psi_R} ) ,
\nonumber\\
&&\Phi_L\ (\ 3,\ 1,\ Y_{\Phi},\  Y'_{\Phi_L} ) ,
\nonumber\\
&&\Phi_R\ (\ 3,\ 1,\ Y_{\Phi},\  Y'_{\Phi_R} ) ,
\nonumber\\
&&\eta_L\ \ (\ 1,\ 1,\  Y_\eta,\  Y'_{\eta_L} ) ,
\nonumber\\
&&\eta_R\ \ (\ 1,\ 1,\  Y_\eta,\  Y'_{\eta_R} ) ,
\label{SpecCase2}
\eea
In this case, the  cancellation conditions for the six anomalies of Eq.(\ref{anomalies}) are given in Eqs.~(\ref{case2}) in the Appendix. We find that there are no non-trivial solutions ($Y'_Q\neq0$) for which $Y_\Phi=n/3$, with $n$ integer. Again, this suggests the existence of stable baryons with fractional electric charge, which is cosmologically unacceptable. So, we consider this possibility unrealistic as well. We have worked out  the complete set of equations (\ref{case2}), finding again 8 possible assignments of charges for the DS of Eq.(\ref{SpecCase2}), which are presented in Table \ref{table:SolutionsCase2} of the Appendix. 

\vspace{0.3cm}
\noindent
In summary, the two minimalistic UV completions, Eqs.(\ref{SpecCase1}, \ref{SpecCase2}),
examined in this section are not phenomenologically viable,
so we have to go a step forward by adding, at least,
one extra $SU(3)_C\times SU(2)_L$ singlet.
This leads to our final minimal scenario, which is discussed in the next section.

\section{The minimal scenario}
\label{sec:minimal}

From the above discussion, the minimal (viable) DS for a leptophobic mediator, $Z'$, axially coupled to quarks, consists of four particles: $\chi_{L,R}, \Phi_{L,R}, \psi_{L,R}, \eta_{L,R}$, with $SU(3)_C\times SU(2)_L \times U(1)_Y \times U(1)_{Y^\prime}$ representations:

\bea
&&\chi_L\ \ (\ 1,\ 1,\  0,\ \ \ Y'_{\chi_L}\ ) ,
\nonumber\\
&&\chi_R\ \ (\ 1,\ 1,\ 0,\ \ \ Y'_{\chi_R}\ ) ,
\nonumber\\
&&\Phi_L\ \ (\ 3,\ 1,\ Y_{\Phi},\  Y'_{\Phi_L}\  ) ,
\nonumber\\
&&\Phi_R\ \ (\ 3,\ 1,\ Y_{\Phi},\  Y'_{\Phi_R}\  ) ,
\nonumber\\
&&\psi_L\ \ (\ 1,\ 2,\ Y_{\psi},\  Y'_{\psi_L}\  ) ,
\nonumber\\
&&\psi_R\ \ (\ 1,\ 2,\ Y_{\psi},\  Y'_{\psi_R}\  ) ,
\nonumber\\
&&\eta_L\ \ \ (\ 1,\ 1,\  Y_\eta,\  Y'_{\eta_L}\ ) ,
\nonumber\\
&&\eta_R\ \ \ (\ 1,\ 1,\  Y_\eta,\  Y'_{\eta_R}\ ) .
\label{MinCase}
\eea

We have assumed here that the $\chi$ particle has vanishing hypercharge, in order to play the role of DM, but the latter could also be played by the neutral component of $\psi$ (if $Y_{\psi}=\pm1/2$).

Now the conditions for the cancellation of the six anomalies of Eq.(\ref{anomalies}) read
\begin{equation}
\begin{split}
\ &12 Y^\prime_Q+(Y^\prime_{\Phi_L}-Y^\prime_{\Phi_R})=0 \\ 
\ &9 Y^\prime_Q+(Y^\prime_{\psi_L}-Y^\prime_{\psi_R})=0 \\ 
\ & \frac{11}{2} Y^\prime_{Q_L}+3 Y_\Phi^2 (Y^\prime_{\Phi_L}-Y^\prime_{\Phi_R})+2Y_\psi^2(Y^\prime_{\psi_L}-Y^\prime_{\psi_R})+Y_\eta^2(Y^\prime_{\eta_L}-Y^\prime_{\eta_R})=0\\ 
\ & 3 Y_\Phi ({Y^\prime_{\Phi_L}}^2-{Y^\prime_{\Phi_R}}^2)+2Y_\psi({Y^\prime_{\psi_L}}^2-{Y^\prime_{\psi_R}}^2)+Y_\eta({Y^\prime_{\eta_L}}^2-{Y^\prime_{\eta_R}}^2)=0\\ 
\ & 36 Y^\prime_{Q_L}+3  (Y^\prime_{\Phi_L}-Y^\prime_{\Phi_R})+2(Y^\prime_{\psi_L}-Y^\prime_{\psi_R})+(Y^\prime_{\eta_L}-Y^\prime_{\eta_R}) +(Y^\prime_{\chi_L}-Y^\prime_{\chi_R})=0 \\ \ & 36 {Y^\prime_{Q_L}}^3+3  ({Y^\prime_{\Phi_L}}^3-{Y^\prime_{\Phi_R}}^3)+2({Y^\prime_{\psi_L}}^3-{Y^\prime_{\psi_R}}^3)+({Y^\prime_{\eta_L}}^3-{Y^\prime_{\eta_R}}^3) + ({Y^\prime_{\chi_L}}^3-{Y^\prime_{\chi_R}}^3)=0 \ .
\end{split}
\label{6eqs}
\end{equation}

This set of equations is difficult to handle. However, it becomes much more tractable by going into a Gr\"obner basis for them \cite{groebner}.
This provides a set of equations, equivalent to (\ref{6eqs}), in which the unknowns can be trivially solved in sequential order, much as in Gaussian elimination for a system of linear equations. Normalizing the extra hypercharges as $Y'_Q=1$, the equivalent set of equations reads

\bea
&&\hspace{-1cm} 2 Y_\eta^2 (Y^\prime_{\chi_L} - Y^\prime_{\chi_R} - 18) + 72 Y_{\Phi}^2 + 36 Y_{\psi}^2 - 11=0 
\label{groeb1}\\
&&\hspace{-1cm} Y^\prime_{\chi_L} - Y^\prime_{\chi_R} + Y^\prime_{\eta_L} - Y^\prime_{\eta_R} - 18 =0\\
&&\hspace{-1cm} {Y_{\chi_L}^\prime}^3 (-A - 72 {Y_\Phi}^2) + {Y_{\chi_R}^\prime}^3 (A + 72 {Y_\Phi}^2) + 
 4 {Y_{\chi_R}^\prime} (-81 (-3 B + 8 C) - 36 {Y_{\eta_R}^\prime} (C - A) + {Y_{\eta_R}^\prime}^2 A)\nonumber \\
&&\hspace{-1cm} + 2 {Y_{\chi_R}^\prime}^2 (-9 (-3 B + 4 C) - {Y_{\eta_R}^\prime} D) - {Y_{\chi_L}^\prime}^2 (-3 {Y_{\chi_R}^\prime} B + 2 (9 (4 C - 3 B) + {Y_{\eta_R}^\prime} D))\nonumber \\
 &&\hspace{-1cm} + 
 {Y_{\chi_L}^\prime} (-3 {Y_{\chi_R}^\prime}^2 B + 4 {Y_{\chi_R}^\prime} (9 (4 C - 3 B) + {Y_{\eta_R}^\prime} D) + 
    4 (81 (8 C - 3 B) - {Y_{\eta_R}^\prime}^2 A + 36 {Y_{\eta_R}^\prime} (C - A))) \nonumber \\
&&\hspace{-1cm}+72 (-18 {Y_{\eta_R}^\prime} (2 C - A) + {Y_{\eta_R}^\prime}^2 A + 
    3 (8 {Y_\Phi}^2 (70 - 27 {Y_{\psi_R}^\prime} + 3 {Y_{\psi_R}^\prime}^2) \nonumber\\
&&\hspace{-1cm}-    3 (99 + 36 C + (-405 + 36 {Y_{\psi_R}^\prime} - 4 {Y_{\psi_R}^\prime}^2) =0 \label{groeb3}\\
&&\hspace{-1cm} Y^\prime_{\psi_L} - Y^\prime_{\psi_R}+9=0\\
&&\hspace{-1cm}({Y_{\chi_R}^\prime} + {Y_{\eta_R}^\prime} + 18) ({Y_{\chi_L}^\prime} - {Y_{\chi_R}^\prime}) ({Y_{\eta_R}^\prime} - {Y_{\chi_L}^\prime} + 18)
 \nonumber \\
&&\hspace{-1cm}+ 18 (2{Y_{\Phi_R}^\prime} ({Y_{\Phi_R}^\prime} - 12) + {Y_{\psi_R}^\prime} ({Y_{\psi_R}^\prime} - 9) - 
    {Y_{\eta_R}^\prime} ({Y_{\eta_R}^\prime} + 18)) + 258=0 \\
&&\hspace{-1cm}  Y'_{\Phi_L} - Y'_{\Phi_R} + 12 =0\ ,
\label{groeb6}
\eea
with
\begin{equation}
\begin{split}
\ & A = -11 + 36 Y_\psi^2 \\
\ &  B = A - 24 Y_\Phi^2\\
\ & C=Y_\eta (-9 + 2 Y_{\psi_R}^\prime) Y_\psi\\
\ &  D= (22 - 72 Y_\psi^2)\ .\\
\end{split}
\label{ABCD}
\end{equation}
%{\roig (I changed equations 17 and 19 (and also 21) since they weren't right and finding the mistakes was impossible for me. These new equations are checked.)}

The free (arbitrary) parameters in the previous equations (\ref{groeb1}-\ref{groeb6}) are
\be
\{Y_\eta,\ Y_\psi,\ Y_\Phi,\ Y'_{\chi_R},\ Y'_{\eta_R}\}\ .
\label{free}
\ee
This means, in particular, that we can choose all the ordinary hypercharges of the DS, so that there are no cosmological problems related to fractional electric charges. Now, each equation in (\ref{groeb1}-\ref{groeb6}) solves one parameter in terms of the precedent ones, so it is trivial, once the initial parameters (\ref{free}) have been chosen, to obtain the others in terms of them. The sequence of reduction goes as
\be
\{Y_\eta,\ Y_\psi,\ Y_\Phi,\ Y'_{\chi_R},\ Y'_{\eta_R}\} \rightarrow Y'_{\chi_L}\rightarrow Y'_{\eta_L}
\rightarrow Y'_{\psi_R}\rightarrow Y'_{\psi_L}\rightarrow Y'_{\Phi_R}\rightarrow Y'_{\Phi_L}\ .
\label{sequence}
\ee
Note that for all the equations the eliminations are linear, and thus completely trivial and unambiguous, except for Eq.(\ref{groeb3}), which is a second-order equation and therefore implies a double solution for $Y'_{\psi_R}$ (and thus for the subsequent variables in the sequence (\ref{sequence})).

Eqs.(\ref{groeb1}-\ref{groeb6}) represent the general solution for the possible hypercharges and extra-hypercharges of the minimal DS (\ref{MinCase}). In order to gain some intuition on the scenario we can particularize the general solution for concrete values of the hypercharges. E.g. for $Y_\eta=1,\ Y_\psi=1/2,\ Y_\Phi=1/3$, we get\footnote{The set of equations (\ref{groebreduced1}-\ref{groebreduced}) is of course equivalent to the set (\ref{groeb1}-\ref{groeb6}) for the above values of $Y_\eta,\ Y_\psi,\ Y_\Phi$. However we have obtained them by replacing those values in the initial equations (\ref{6eqs}) and then going into a Gr\"obner basis.} 
%{\roig (I changed the equations for the new case with $Y_\Phi^\prime=1/3$)}

\bea
&&  Y'_{\chi_L} - Y'_{\chi_R} - 15 = 0
\label{groebreduced1}
\\
&&- 3 + Y'_{\eta_L} - Y'_{\eta_R}=0\\
&&371 - 300 Y'_{\chi_R} - 20 {Y'_{\chi_R}}^2 + 78 Y'_{\eta_R} \nonumber \\
&&- {Y'_{\eta_R}}^2 - 486 Y'_{\psi_R} - 18 Y'_{\eta_R} Y'_{\psi_R} + 51 {Y'_{\psi_R}}^2=0 \\
&&9 + Y'_{\psi_L} - Y'_{\psi_R}=0\\
&&-39 -  Y'_{\eta_R}+4 Y'_{\Phi_R}+3 Y'_{\psi_R}=0 \\
&&-9 - Y'_{\eta_R}+4 Y'_{\Phi_L}+3 Y'_{\psi_R}=0\ ,
\label{groebreduced}
\eea
with the same sequence of reduction as (\ref{sequence}). Some particular solutions to Eqs.(\ref{groebreduced1}-\ref{groebreduced}) are shown in  Table \ref{particular_examples}.

\begin{table}[htbp]
\begin{center}
\begin{tabular}{| c | c | c | c | c | c | c | c | c | }
\hline
%$Y_{\eta}$ & 1 & 1 & 1 & 1 & 1 & 1   \\
%\hline
%$Y_{\psi}$ & 1/2 & 1/2 & 1/2 & 1/2 & 1/2 & 1/2  \\
%\hline
%$Y_{\Phi}$ & 1/3 & 1/3 & 1/3 & 1/3 & 1/3 & 1/3   \\
%\hline
$Y_{\chi_R}^\prime$ & 1 & 1 & 1 & 1 & 2 & 2  \\
\hline 
$Y_{\eta_R}^\prime$ & 1 & 1 & 2 & 2 & 1 & 1  \\
\hline
$Y_{\chi_L}^\prime$ & 16 & 16 & 16 & 16 & 17 & 17   \\
\hline
$Y_{\eta_L}^\prime$ & 4 & 4 & 5 & 5 & 4 & 4  \\
\hline
$Y_{\psi_R}^\prime$ & 0.260 & 9.621 & 0.404 & 9.830 & -0.440 & 10.323  \\
\hline
$Y_{\psi_L}^\prime$ & -8.739 & 0.621 & -8.595 & 0.830 & -9.440 & 1.323  \\
\hline
$Y_{\Phi_R}^\prime$ & 9.804 & 2.783 & 9.946 & 2.877 & 10.330 & 2.257 \\
\hline
$Y_{\Phi_L}^\prime$ & 13.045 & 10.705 & 13.982 & 11.625 & 13.221 & 10.530  \\
\hline 
\end{tabular}
\caption{Explicit examples of extra-hypercharge assignments in the minimal DS, Eq.(\ref{MinCase}), that lead to anomaly cancellation. (For the non-rational charges, only the first decimals are shown.) The extra-hypercharges of the SM quarks, Eq.(\ref{SMcharges}), are normalized as $Y'_Q=1$. The ordinary hypercharges of the DS fermions are $Y_{\eta}=1$, $Y_{\psi}=1/2$ and $Y_{\Phi}=1/3$.}
\label{particular_examples}
\end{center}
\end{table}

\vspace{0.3cm}
%\noindent
One could try to get additional examples of UV completion by going beyond the minimal DS studied in this paper. One possibility, examined in Ref.~\cite{Ismail:2016tod}, is to consider a DS consisting of a whole SM-like vectorial family. In this way, a model (named Model 4) was obtained in \cite{Ismail:2016tod} with rational (though still weird) extra-hypercharges. 

Another (even less economical, but somehow trivial) solution is to assign to every SM fermion a DS fermion with the same representation and charges, but opposite chirality. In this way all fermions form vectorial pairs and anomaly cancellation is automatic. This obvious possibility was also noticed in Ref.~\cite{Ismail:2016tod}. Then the DM fermion, $\chi_{L,R}$, must correspond to a couple of right-handed neutrinos with non-vanishing extra-hypercharge. Besides, the Higgs sector must be further extended to incorporate  Yukawa couplings for both charged and neutral leptons. Notice that in a scenario of this kind, the remarkable anomaly cancellation inside the ordinary SM would be a (weird) accident.

\section{Phenomenological perspective}
\label{sec:pheno}
In this section we examine the main differences between the simplified model approaches (e.g. \cite{Lebedev:2014bba,
Hooper:2014fda,Bagnaschi:2019djj})
and the UV-complete scenario of an axial, leptophobic $Z'$
described by the most minimal scenario of Sec.\ref{sec:minimal}.
%\footnote{We note from Table~\ref{particular_examples} that at least some of the exotics tend to have irrational $U(1)_{Y^\prime}$ charges which, although not forbidden in principle, can be theoretically disfavoured\cite{Li:1981un,Banks:2010zn}.}.
\begin{description}
\item[Kinetic mixing]

The presence of an extra $U(1)$ interaction opens the door to a dangerous kinetic mixing between the standard $B-$boson and the one associated to $U(1)_{Y'}$,
\be
{\cal L}_{\rm kin}\supset -\frac{1}{2}\ \epsilon\  F^Y_{\mu\nu}{F^{Y'}}^{\mu\nu}\ .
\ee
This mixing contributes to the $S$ and $T$ parameters and, most importantly, to dilepton production at the LHC. One can set $\epsilon=0$ at some scale $\Lambda$ (presumably the scale of symmetry breaking of a unifying gauge group), but still the mixing is radiatively generated through loops involving particles with non-vanishing $Y, Y'$ charges. In the case of a $Z'$ with vectorial coupling to quarks, the contribution of the latter is $\Delta \epsilon\simeq 0.02\ g_{Y'}Y'_Q\ \log\Lambda/\mu$, where $\mu\sim m_{Z'}$ \cite{Kahlhoefer:2015bea}. This translates in bounds on the gauge coupling. E.g. for $\Lambda=10$ TeV and $m_{Z'}= 200$ GeV (1 TeV) one gets $g_{Y'}Y'_Q < 0.1$ (1) \cite{Kahlhoefer:2015bea}. As it was pointed out in Ref.~\cite{Kahlhoefer:2015bea}, in the case of axial coupling the quarks do not contribute to the mixing, since their contributions cancel as a consequence of ${\rm Tr}\ Y=0$ in the quark sector, see Eq.(\ref{SMcharges}). However, the dark leptons, whose presence is obliged in the UV completion, do contribute to the mixing, namely
\bea
\Delta \epsilon &=& \frac{e g_{Y'}}{12\pi^2\cos \theta_W}\ \left[2Y_\psi(Y'_{\psi_L}-Y'_{\psi_R})+3Y_\Phi(Y'_{\Phi_L}-Y'_{\Phi_R})+(Y'_{\eta_L}-Y'_{\eta_R})\right]\ \log\Lambda/\mu
\nonumber\\
&\simeq& 0.003\ {\cal O}(10)\ g_{Y'} \log\Lambda/\mu\ ,
\label{kinmix}
\eea
where we have used the values of the charges of Table 1. Note that the extra hypercharges, $Y'_{\psi_L}, Y'_{\psi_R}$, etc. depend on the model but they are always ${\cal O}$(10) due to the anomaly cancellation conditions. Consequently, in contrast to previous simplified analyses, for the axial case it continues to be true that a kinetic mixing is generated with a similar size as in the vectorial instance. It is also worth-noticing that the presence of two Higgs doublets, $H_u, H_d$ with non-vanishing $Y, Y'$ charges (see Sec. \ref{sec:higgssector}) does potentially contribute to $\epsilon$. However, the fact that they possess the same (opposite) $Y$ ($Y'$) charge makes their contributions to cancel. 

\item[Mass mixing]
The presence of the two Higgs doublets, $H_u, H_d$, with non-vanishing $Y, Y'$ charges does lead however to a mixing term in the $Z'-B$ mass matrix, $\sim g'g_{Y'}(v_d^2 - v_u^2)$; and thus to a contribution to the $Z-Z'$ mixing angle. More precisely, in the absence of kinetic mixing, denoting the gauge eigenstates as $\hat Z$ and $\hat Z'$, this is given by 
\begin{equation}
\tan 2 \theta'  =  \frac{ 2 g_{Y'} (v_d^2 - v_u^2) (2 Y^\prime_{Q}) g' }{  {g}_{Y'}^2 (v_d^2 + v_u^2) (2 Y^\prime_{Q})^2 + {m}_{\hat Z'}^2 - {m}_{\hat Z}^2}.
\label{thetaprime}
\ee
where $\langle H_u \rangle = \frac{1}{\sqrt 2}(0,v_u)$, $\langle H_d \rangle = \frac{1}{\sqrt 2}(v_d,0)$, $\langle H_l \rangle = \frac{1}{\sqrt 2}(v_l,0)$, with $v^2 = v_u^2 + v_d^2 +v_l^2  = (246\, \GeV)^2$, and  $Y'_{H_u} = Y'_{H_d} = 2 Y^\prime_{Q}$.
In the limit ${m}_{\hat Z'} \gg {m}_{\hat Z}$, the above equation reads
\begin{equation}
\theta' \simeq \frac{g_{Y'} (v_d^2 - v_u^2) (2 Y^\prime_{Q}) g'}{{m}_{\hat Z'}^2}.
\label{thetaprimeaprox}
\end{equation}
It should be noticed that this source of mixing is totally model-independent, since the presence of the two Higgs states in the quark sector with non-vanishing $Y, Y'$ charges is a direct consequence of the axial coupling (see Sec.\ref{sec:higgssector}). As a matter of fact, this contribution to the mixing can be more important than that from the previous kinetic mixing. Current limits on $\theta'$ come from electroweak precision tests~\cite{Erler:2009jh,delAguila:2010mx} and $WW$~\cite{Pankov:2019yzr,Bobovnikov:2018fwt,Aad:2019fbh,Sirunyan:2017acf}/dilepton~\cite{Aad:2019fac,CMS-PAS-EXO-19-019} production at the LHC. Typically the bounds are at the per mil level (although the strongest bounds, coming from LHC, are somewhat model-dependent). Hence, we expect  lower bounds on the $Z'$ mass of the order
\begin{equation}
m_{Z'} \simgt 26 \sqrt{ g' g_{Y'}Y^\prime_{Q}\cos 2\beta}\  \tilde v ,
\label{mzpbound}
\end{equation}
where $\tilde v^2=v_u^2 + v_d^2$ and $v_u = \tilde v \sin \beta$, $v_d = \tilde v \cos \beta$. We must note that $\tilde v$ cannot be much smaller than $v$, otherwise the top Yukawa coupling would become non-perturbative.

\item[Perturbativity limits]
The large $Y'$ charges in the minimal dark sector, required for anomaly cancellation, impose perturbativity limits on $g_{Y'}$. Although the particular values depend upon the model (see Table \ref{particular_examples}), there exist some regularities. In particular, the largest difference between the left and right extra hypercharges corresponds to the dark matter: $Y'_{\psi_L}- Y'_{\psi_R}=15$, which suggests the presence of a scalar with charge $Y'_S=15$  to provide mass to the dark matter. Then, perturbativity demands $g_{Y'}\simlt {\cal O}(10^{-1})$.

\item[Extra scalars]
The phenomenological viability of the minimal dark sector requires the presence of several extra scalar fields. First of all, there must be one (or several) scalar(s) responsible for the $U(1)_{Y'}$ breaking. Certainly, the obliged presence of at least three Higgs states, $H_u, H_d, H_l$ allows in principle to give mass to the $Z'$ without any extra scalar state. However, this would imply unacceptably large $Z-Z'$ mixings unless the coupling becomes negligibly small. Consequently, one extra scalar, $S$, is required to play the dominant role in the $U(1)_{Y'}$ breaking. Besides, its VEV (times $Y'_S$) must be much larger than those of the Higgses, to avoid too large off-diagonal entries in the $Z-Z'$ mass matrix, as discussed above.

The mass of this scalar is restricted by unitarity, see Ref.~\cite{Kahlhoefer:2015bea},
 \begin{equation}
   m_{S} \lesssim \frac{\pi m_{Z'}^2}{{g^A_{Y^\prime}}^2 m_f},
   \label{msbound}
 \end{equation}
 where $f$ is the heaviest fermion in the theory with axial coupling, $g^A_f$, to $Z'$. In particular $g^A_{f}=g_{Y^\prime}Y'_Q$ for the top quark and $g^A_f=g_{Y^\prime}(Y'_{f_R}-Y'_{f_L})/2$ for the dark fermions.

Notice also that scalar fields are required to give masses to the dark fermions: $\chi, \psi, \eta, \Phi$. The fact that the $Y'$ charges are not the same implies the presence of at least 4 extra scalars in the model.

\item [Dark matter annihilation]
  The most obvious modification of DM phenomenology induced by the UV completion of the model is the role played by the extra scalar (S) and the dark fermions in DM annihilation. The first issue was considered in Ref.~\cite{Kahlhoefer:2015bea}, but the second was not. More precisely, the presence of extra fermions with non-trivial representation under the SM gauge group can induce co-annihilation effects if their masses are not far from the DM one. As a consequence the required value of $g_{Y'}$ to achieve the correct relic density might be relaxed.
  
\end{description}

\section{Conclusions}
\label{sec:conc}
The $Z'$-portal is one of most popular and well-founded scenarios of dark matter (DM). However, it is  subject to severe experimental and observational constraints, in particular those coming from di-lepton production at the LHC and from DM direct-detection experiments. Consequently, it is often required that the $Z'$ boson has couplings which are (i) leptophobic, (ii) axial either with the DM particle or with
the quarks (or both). Condition (ii) leads to spin-dependent direct-detection cross-section, maybe with velocity suppression. 

Most of the analyses so far have been performed in the context of simplified models. However, it is convenient to consider their possible UV completions, not only for the sake of theoretical consistency but also from the phenomenological point of view. It turns out that, e.g. the requirement of anomaly cancellation implies the existence of an extended dark sector (beyond a lone DM particle) and strong correlations between the $U(1)'$ charges of the SM and the dark sector (DS) fermions. This is of great importance for phenomenological analyses, as well as for evaluations of the relic density.

Concerning UV completions, the case in which the leptophobic $Z'$ has axial couplings to the DM has been well studied in the current literature. However, the complementary case, when the $Z'$ presents axial couplings to the quarks is still essentially unexplored (except for Ref.~\cite{Ismail:2016tod}).

In this paper we have considered the latter scenario, building up the minimal DS (from the point of view of the spectrum) that is anomaly-free and contains a candidate for DM particle. It turns out that the most economical possibilities are not phenomenologically viable since they contain fractional electrically-charged particles. Then, the minimal DS consists of four particles: a SM singlet (the DM particle, $\chi_{L,R}$), a $SU(3)_C$ triplet ($\Phi_{L,R}$), a $SU(2)_L$ doublet ($\psi_{L,R}$) and a  $SU(3)_C\times SU(2)_L$ singlet ($\eta_{L,R}$) (see Eq.(\ref{MinCase})). This means, in particular, that the minimal DS is larger than the analogous one when the $Z'$ has vectorial coupling to quarks.

Regarding the possible assignments of (ordinary and extra) charges to the various fields, the complete set of solutions to the anomaly-cancellation conditions can be expressed in a convenient form using a Gr\"obner basis, as explained in Eqs.(\ref{groeb1}-\ref{sequence}). It turns out, in particular, that it
is possible to choose the hypercharges of the DS fields, so that no fractional electric-charge states are present. Then the consistency equations become simpler. Still, the set of solutions contains two free parameters, which we have chosen as $Y'_{\chi_R}, Y'_{\eta_R}$, as indicated in Eq.(\ref{sequence}). However, the solutions imply the existence of non-rational $U(1)'$ charges. Some examples are given in Table \ref{particular_examples}. Going beyond the minimal DS it is possible to get rational (but still weird) charges.

Concerning the Higgs sector, we have shown that the consistency of Yukawa couplings requires at least three Higgs states giving mass to $u-$quarks, $d-$quarks and charged leptons, respectively. This result holds for any consistent (minimal or not) DS.

We have also described the key features of the phenomenology of the UV complete scenario as compared to its simplified-model counterparts. We noticed that, unlike the quarks, the extra (dark) fermions do contribute to the kinetic $Z-Z'$ mixing, in an amount similar to that of vectorial models, see Eq.(\ref{kinmix}). In addition, the presence of multiple Higgs bosons charged under $U(1)_{Y'}$ gives rise to a significant and model-independent contribution to the $Z-Z'$ mass mixing, see Eqs.(\ref{thetaprime}, \ref{thetaprimeaprox}) . Then, the existence of stringent experimental bounds on the $Z-Z'$ mixing angle can be translated into a lower limit on $m_Z'$ as a function of the Higgs VEVs, see Eq.(\ref{mzpbound}). Moreover, the existence of several (at least 4) extra singlet scalar fields are guaranteed from the requirement of generating masses of the dark fermions. If not too heavy, these can induce notable changes in the phenomenology of the model. In particular, the extra scalars and fermions can help the DM to achieve correct relic density with relatively smaller values of the coupling.

\section*{Acknowledgments}

This work has been partially supported by MINECO, Spain, under contract 
%FPA2014- 57816-P, 
FPA2016-78022-P and Centro de excelencia Severo Ochoa Program under grant SEV-2016-0597. We thank J.M. Moreno for illuminating discussions concerning the Gr\"obner basis.

\appendix
\newpage

\section{Appendix}
\label{app}

In this appendix we summarize the charge assignments for the smallest UV completions of an axial, leptophobic $Z'$ model, which fulfil anomaly cancellation. These were discussed in Sec.\ref{sec:anomaly}, where it was stressed that these ``minimalistic" solutions have the shortcoming of involving particles with fractional electric charges, so they are not phenomenologically viable (the minimal phenomenologically acceptable solution is discussed in Sec.\ref{sec:minimal}).

One set of such solutions corresponds to the matter content of the dark sector as given by Eq.(\ref{SpecCase1}). 
The corresponding cancellation conditions for the six anomalies of Eq.(\ref{anomalies}) read
\begin{equation}
\begin{split}
\ &12 Y^\prime_Q+(Y^\prime_{\Phi_L}-Y^\prime_{\Phi_R})=0 \\
\ &9 Y^\prime_Q+(Y^\prime_{\psi_L}-Y^\prime_{\psi_R})=0 \\
\ & \frac{11}{2} Y^\prime_{Q} +3 Y_\Phi^2 (Y^\prime_{\Phi_L}-Y^\prime_{\Phi_R})+2Y_\psi^2(Y^\prime_{\psi_L}-Y^\prime_{\psi_R})=0\\
\ & 3 Y_\Phi ({Y^\prime_{\Phi_L}}^2-{Y^\prime_{\Phi_R}}^2)+2Y_\psi({Y^\prime_{\psi_L}}^2-{Y^\prime_{\psi_R}}^2)=0  \\
\ &36 {Y_Q^\prime}^3+3({Y_{\Phi_L}^\prime}^3-{Y_{\Phi_R}^\prime}^3)+2({Y_{\psi_L}^\prime}^3-{Y_{\psi_R}^\prime}^3)+({Y_{\chi_L}^\prime}^3-{Y_{\chi_R}^\prime}^3)=0 \\
\ &36 {Y_Q^\prime}+3({Y_{\Phi_L}^\prime}-{Y_{\Phi_R}^\prime})+2({Y_{\psi_L}^\prime}-{Y_{\psi_R}^\prime})+({Y_{\chi_L}^\prime}-{Y_{\chi_R}^\prime})=0  
\end{split}
\label{case1}
\end{equation}
%
%{\roig (equations completed)}
It is worth-noticing that the first three equations imply
\be
Y^\prime_{Q} (72 Y_\Phi^2+36 Y_\psi^2-11)=0 \ ,
\ee
which only has non-trivial solution ($Y'_Q\neq 0$) if $Y_\psi=\pm1/2$, $Y_{\Phi}=\pm 1/6$.
Solving the complete set of equations (\ref{case1}) we find the 8 possible
assignments of charges for the DS presented in Table \ref{table:SolutionsCase1}.
Note there that all charges are given in terms of two parameters, $Y'_Q$ and $Y'_{\psi_R}$,
which are arbitrary. Furthermore, as mentioned above, there is a trivial factor
of proportionality for all $Y'-$charges, so we have taken $Y'_Q=1$ with no loss of generality.

\vspace{0.2cm}

\begin{table}[htbp]
\begin{center}
\begin{tabular}{| c | c | c | c | c | c | c | c | c | }
\hline
$Y_{\psi}$ & 1/2 & 1/2 & -1/2 & -1/2  \\
\hline
$Y_{\Phi}$ & 1/6 & -1/6 & 1/6 & -1/6  \\
\hline
$Y_{\Phi_R}^\prime$ & $\frac{3}{4}( {17 }- 2Y_{\psi_R}^\prime )$ & $-\frac{3}{4}( {1}- 2Y_{\psi_R}^\prime )$ & $-\frac{3}{4}( {1}- 2Y_{\psi_R}^\prime )$ & $\frac{3}{4}( {17}- 2Y_{\psi_R}^\prime )$  \\
\hline 
$Y_{\Phi_L}^\prime$ &  $\frac{3}{4}( 1- 2 Y_{\psi_R}^\prime )$ & $-\frac{3}{4}( 17 - 2Y_{\psi_R}^\prime )$ & $-\frac{3}{4}( 17 - 2Y_{\psi_R}^\prime )$ & $\frac{3}{4}( 1- 2 Y_{\psi_R}^\prime )$   \\
\hline
$Y_{\psi_L}^\prime$ & \multicolumn{4}{c|}{$Y_{\psi_R}^\prime-9 $} \\
\hline
$Y_{\chi_R}^\prime$ &  \multicolumn{4}{c|}{$-9 \pm\frac{1}{2}\sqrt{22 Y_{\psi_R}^{\prime 2}-198 Y_{\psi_R}^{\prime} +2747/6}$} \\
\hline
$Y_{\chi_L}^\prime$ &  \multicolumn{4}{c|}{$9  \pm\frac{1}{2}\sqrt{22 Y_{\psi_R}^{\prime 2}-198 Y_{\psi_R}^{\prime}+2747/6}$}   \\
\hline 
\end{tabular}
\renewcommand{\thetable}{A.1}
\caption{Charge assignments for the DS of Eq.\ref{SpecCase1} satisfying anomally cancellation conditions of Eq.\ref{case1}. $Y_{\psi_R}^{\prime}$ is a free parameter. The expressions correspond to the normalization $Y_{Q}^{\prime}=1$. In general, one should understand $Y_f^\prime$ above as $Y_{f}^{\prime}/Y_{Q}^{\prime}$ for all fermions $f$
(including $Y_{\psi_R}^{\prime}$ inside the expressions). The two $\pm$ signs are correlated, so for each value of $Y_{\psi_R}^{\prime}$ there are 8 solutions.}

\label{table:SolutionsCase1}
\end{center}
\end{table}

As discussed in Sec.\ref{sec:anomaly}, the fact that the dark quarks ($\Phi$) have
electric charge $Q^{\rm el}_\Phi=\pm1/6$ strongly suggests the existence of
stable baryons with fractional electric charge, with disastrous cosmological consequences.
Thus, we consider this possibility unrealistic.

\vspace{0.3cm}
\noindent
The other set of such solutions corresponds to the dark sector of Eq.(\ref{SpecCase2}), with $Y_\psi=\pm 1/2$ to enable a DM particle. In this case, the cancellation conditions for the six anomalies of Eq.(\ref{anomalies}) read

\begin{equation}
\begin{split}
\ &12 Y^\prime_Q+(Y^\prime_{\Phi_L}-Y^\prime_{\Phi_R})=0 \\
\ &9 Y^\prime_Q+(Y^\prime_{\psi_L}-Y^\prime_{\psi_R})=0 \\
\ & \frac{11}{2} Y^\prime_{Q}+3 Y_\Phi^2 (Y^\prime_{\Phi_L}-Y^\prime_{\Phi_R})+2Y_\psi^2(Y^\prime_{\psi_L}-Y^\prime_{\psi_R})+Y_\eta^2(Y^\prime_{\eta_L}-Y^\prime_{\eta_R})=0\\
\ & 3 Y_\Phi ({Y^\prime_{\Phi_L}}^2-{Y^\prime_{\Phi_R}}^2)+2Y_\psi({Y^\prime_{\psi_L}}^2-{Y^\prime_{\psi_R}}^2)+Y_\eta({Y^\prime_{\eta_L}}^2-{Y^\prime_{\eta_R}}^2)=0\\
\ & 36 {Y^\prime_{Q}}^3+3  ({Y^\prime_{\Phi_L}}^3-{Y^\prime_{\Phi_R}}^3)+2({Y^\prime_{\psi_L}}^3-{Y^\prime_{\psi_R}}^3)+({Y^\prime_{\eta_L}}^3-{Y^\prime_{\eta_R}}^3)=0 \\
\ &  36 Y^\prime_{Q}+3  (Y^\prime_{\Phi_L}-Y^\prime_{\Phi_R})+2(Y^\prime_{\psi_L}-Y^\prime_{\psi_R})+(Y^\prime_{\eta_L}-Y^\prime_{\eta_R})=0 \ .
\end{split}
\label{case2}
\end{equation}

%{\roig (equations completed)}
It is straightforward to check that the first five equations lead to
\be
Y^\prime_Q (72 Y_\Phi^2+36 Y_\psi^2-36 Y_{\eta}^2-11)=0\ .
\ee
Keeping in mind that $Y_\psi=\pm 1/2$, it is easy to see that this equation does not have non-trivial solutions ($Y'_Q\neq0$) for which $Y_\Phi=n/3$, with $n$ integer. Again, this suggests the existence of stable baryons with fractional electric charge, which is cosmologically unacceptable. So, we consider this possibility unrealistic as well.

Nevertheless, we have found the complete set of solutions, namely the 
8 possible assignments of charges presented in Table \ref{table:SolutionsCase2}.

\begin{table}[htbp]
\begin{center}
\begin{tabular}{| c | c | c | c | c | c | c | c | c | }
\hline
$Y_{\psi}$ & 1/2 & -1/2  \\
\hline
$Y_{\Phi}$ & $\pm\frac{1}{6}\sqrt{18 Y_\eta^2+1}$ & $\pm\frac{1}{6}\sqrt{18 Y_\eta^2+1}$   \\
\hline
$Y_{\psi_L}^\prime$ & $Y_{\psi_R}^\prime-9 $ & $Y_{\psi_R}^\prime-9 $   \\
\hline
$Y_{\eta_R}^\prime$ & $\pm\Sigma-9 Y_\eta Y_{\psi_R}^\prime + (81Y_\eta/2-9)$ & $\pm\Sigma-9 Y_\eta Y_{\psi_R}^\prime + (81Y_\eta/2-9)$  \\
\hline
$Y_{\eta_L}^\prime$ & $\frac{1}{2 Y_\eta^2} \left(Y_\eta^2 Y_{\eta_R}^\prime+72 (Y_\Phi^2-2)\right)$ & 
$\frac{1}{2 Y_\eta^2} \left(Y_\eta^2 Y_{\eta_R}^\prime+72 (Y_\Phi^2-2)\right)$  \\
\hline 
$Y_{\Phi_R}^\prime$ & $\frac{1}{72 Y_\Phi}\left(Y_\eta(Y_{\eta_L}^{\prime 2}-Y_{\eta_R}^{\prime 2})+9((48 Y_\Phi+9) -2Y_{\psi_R}^\prime)\right)$ & $\frac{1}{72 Y_\Phi}\left({Y_\eta(Y_{\eta_L}^{\prime 2}-Y_{\eta_R}^{\prime 2})+9((48 Y_\Phi-9) +2Y_{\psi_R}^\prime)}\right)$  \\
\hline 
$Y_{\Phi_L}^\prime$ & $Y_{\Phi_R}^\prime-12 $ & $Y_{\Phi_R}^\prime-12 $  \\
\hline
\end{tabular}
\renewcommand{\thetable}{A.2}
\caption{Charge assignments for the DS of Eq.\ref{SpecCase2} satisfying anomally cancellation conditions of Eq.\ref{case2}, with $\Sigma=\frac{1}{2\sqrt{6}}\sqrt{(18 Y_\eta^2+1)(132 Y_{\psi_R}^{\prime 2}-1188Y_{\psi_R}^\prime +2747)}$. The free parameters are $Y_{\eta}$, $Y_{\psi_R}^\prime$. For each choice of them, the remaining charges are obtained recursively following the order of the table. In each column the $\pm$ signs are not correlated, thus leading to 8 solutions in total. As for Table \ref{table:SolutionsCase1}
the normalization $Y_{Q}^{\prime}=1$ has been assumed.
}
\label{table:SolutionsCase2}
\end{center}
\end{table}

\bibliography{main}

\end{document}